\documentclass{PoS}

\title{Nucleosynthesis Calculations from Core-Collapse Supernovae}

\ShortTitle{Supernova Nucleosynsthesis}

\author{
\speaker{Christopher L. Fryer$^{ab}$}, 
Patrick Young$^{ac}$, 
Michael Bennett$^{ad}$,
Steven Diehl$^{abe}$,
Falk Herwig$^{adg}$,
Raphael Hirschi$^{ad}$,
Aimee Hungerford$^{ab}$,
Marco Pignatari$^{adg}$,
Georgios Magkotsios$^{acg}$,
Gabriel Rockefeller$^{ab}$,
and Francis X. Timmes$^{ac}$ \\
\llap{$^a$}The NuGrid Collaboration\\
\llap{$^b$}Computational Methods (CCS-2), Los Alamos National Laboratory, Los Alamos, NM, 87544, USA\\
\llap{$^c$}School of Earth and Space Exploration, Arizona State University, Tempe, AZ 85287, USA\\
\llap{$^d$}Astrophysics Group, Keele University, ST5 5BG, UK\\
\llap{$^e$}Theoretical Astrophysics Group (T-6), Los Alamos National Laboratory, Los Alamos, NM, 87544, USA\\
\llap{$^f$}Dept. of Physics \& Astronomy, Victoria, BC, V8W 3P6, Canada\\
\llap{$^g$} Joint Institute for Nuclear Astrophysics, University of Notre Dame, IN, 46556, USA\\
E-mail:\email{fryer@lanl.gov}
}

\abstract{We review some of the uncertainties in calculating
nucleosynthetic yields, focusing on the explosion mechanism.  Current
yield calculations tend to either use a piston, energy injection, or
enhancement of neutrino opacities to drive an explosion.  We show that
the energy injection, or more accurately, an entropy injection
mechanism is best-suited to mimic our current understanding of the
convection-enhanced supernova engine.  The enhanced neutrino-opacity
technique is in qualitative disagreement with simulations of
core-collapse supernovae and will likely produce errors in the yields.
But piston-driven explosions are the most discrepant.  Piston-driven
explosion severely underestimate the amount of fallback, leading to
order-of-magnitude errors in the yields of heavy elements.  To obtain
yields accurate to the factor of a few level, we must use entropy or
energy injection and this has become the NuGrid collaboration
approach.}

\FullConference{10th Symposium on Nuclei in the Cosmos\\
		 July 27 - August 1 2008\\
		 Mackinac Island, Michigan, USA}

\begin{document}

\section{Nucleosynthesis and Understanding Supernova Explosions}

The first step in producing a yield for core-collapse supernovae is to
introduce a realistic explosion.  Although scientists are still
working hard to determine the exact physics behind core-collapse
supernova explosions, there is growing support for the
convection-driven mechanism~\cite{Her94,Fry02,Bur06,Burrows06}.  Even
so, a lot of work remains (both in the progenitor evolution and the
explosion mechanism itself) if we want to accurately predict the
explosion energy for a given stellar mass.  For the foreseeable
future, we will have to artificially induce explosions and explore a
range of explosion energies (producing error bars) for nuclear yields.
However, a qualitative understanding of the explosion mechanism can
help us better induce these explosions so that our range of answers
will actually bracket the true answer.  As we shall see, some
mechanisms used to induce explosions will not produce results
consistent with the convection-enhanced neutrino driven mechanism.

\begin{figure}
\includegraphics[width=0.98\textwidth]{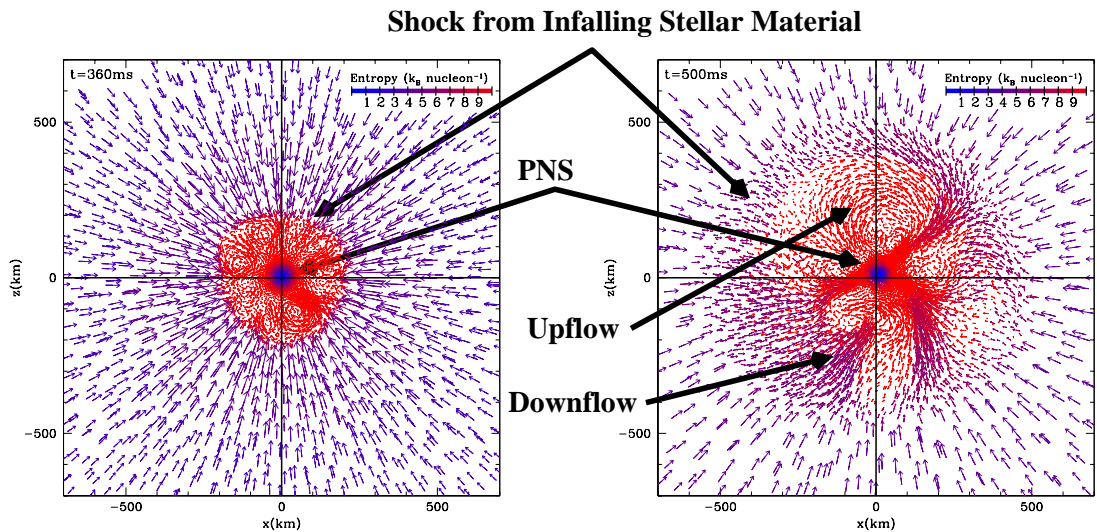}
\caption{A slice of the x-y plane of a 3-dimensional supernova
explosion calculation modeling the collapse of a 23\,M$_\odot$
star~\cite{FY07}.  The proto-neutron star (PNS) and outer edge of the
convective region defined at the position where the infalling stellar
material shocks against the convection are labeled.  Note that this 
outer edge moves outward with time.  Energy is injected into this 
convective region rather uniformly, striving to produce a constant 
entropy profile as the region expands.}
\label{fig:expl}
\end{figure}

Our current understanding of the explosion mechanism behind
core-collapse supernovae involves a series of phases (for a review,
see~\cite{Fry03}).  When the mass of a core becomes so large that
electron capture and iron dissociation can occur, the core collapses.
The core collapses to nuclear densities and bounces, sending a shock
through the star.  Most of the energy in the shock is thermal and,
when neutrinos can escape the shock, they sap the bounce shock's
energy and itself.  After the bounce shock stalls, the region between
the edge of the proto-neutron star and the shock front of the
infalling material is unstable to a number of convective
instabilities.  It is in this convective region that neutrino energy
leaking out of the core is converted into kinetic energy that
eventually pushes out the infalling star and drives an explosion.
Figure~\ref{fig:expl} shows annotated plots of a convection-driven
explosion at 2 different times.  As the energy in the convective
region grows (and the infall rate decreases), the outer edge of the
convective region moves out.  Ultimately, the convective region has
enough energy to drive an explosion (although in this case, the
explosion is so weak that most of the star will fall back onto the
proto-neutron star, forming a black hole).

What can we learn from simulations of this convection-enhanced
explosion mechanism?  First, energy is deposited in a region covering
a few tenths of a solar mass.  Convection strives to flatten the
entropy gradient, so energy is deposited fairly uniformly across the
convective region.  As long as the shock (outer edge of the convective
region) is moving out slowly (slow enough that convection can
redistribute the energy), energy is deposited throughout the
convective region.  In mass coordinates, the region does not change
dramatically with time.  Finally, no energy is deposited beyond the
convective region.

With this understanding, let's compare the different mechanisms
currently used to drive explosions for nucleosynthesis: piston-driven
explosions, energy-driven explosions, and enhanced neutrino-opacity
driven explosions.  Piston-driven explosions have been used extensively 
in the past and much of the comprehensive yields in the literature are 
based on these explosions.  Piston-driven explosions work by placing a hard
surface at the inner boundary (generally assumed to be at the edge of
the iron core, but it would be more realistic to use the outer edge of
the convective region).  This hard surface is then pushed outward,
accelerating the star and driving an explosion.  Such an approximation
keeps accelerating the inner material in the ejecta, not allowing it
to slow down and ultimately fall back on the star, severely
underestimating the fallback and overestimating the amount of heavy 
elements (such as $^{56}$Ni) for a given explosion energy.  This has 
been discussed at some detail~\cite{YF07} and it is now generally 
accepted in the explosion community that piston-driven yields are 
not accurate.

Two alternate options are being used in the literature.  A straight
energy deposition in the inner few tenths of a solar mass.  This
method is designed to incorporate the energy increase in the
convective region.  The energy is limited to a few tenths of a solar
mass (the rough mass size of the convective region throughout most
calculations).  And the energy is injected uniformly (in specific
internal energy) across this region (as we would expect from
convection).  More realistic might be to uniformly increase the
entropy throughout this region.

The second method is to artificially increase the neutrino opacity in
this region.  The argument for this method is that it includes
neutrino changes to the electron fraction (albeit at an exaggerated
level).  The disadvantages are many.  First, it injects energy even
beyond the convective region.  Although it is true that neutrinos in a
realistic engine will do this, the opacity is lower, so we are
over-estimating this energy injection.  Second, the energy deposition
is highly peaked toward the dense material and not distributed across
the convective engine as we would expect in a real convection-enhanced
supernova.  Although these artifacts are probably small when compared
to the piston/energy deposition differences, they all point a
direction opposite from what we would expect from the convection
mechanism.  This method will be less like the convective engine than a
simple direct energy deposition.

\begin{table}
\begin{tabular}{|l|l|l|c|c|c|c|c|}
\hline
Model Name & \multicolumn{2}{c|}{Model Charact.} & \multicolumn{5}{c|}{Yields} \\
and  & $E_{\rm exp}$ & $M_{\rm rem}$ & $^{28}$Si & $^{45}$Sc & $^{44}$Ti & $^{60}$Co & $^{56}$Ni \\
citation & $10^{51}$\,erg & $M_\odot$ & M$_\odot$ & $10^{-5}$\,M$_\odot$ & $10^{-5}$\,M$_\odot$ & $10^{-5}$\,M$_\odot$ & M$_\odot$ \\
\hline
\hline
WW-S22A\cite{WW95} & 1.47 & 2.02 & 0.356 & 1.20 & 6.15 & $2.43$ & 0.205 \\
WW-S25A\cite{WW95} & 1.18 & 2.07 & 0.315 & 0.228 & 3.04 & 5.36 & 0.129 \\
23e-1.5\cite{YF07} & 3.2 & 1.5 & 0.303 & 0.082 & 0.513 & 1.03 & 0.0013 \\
23e-2.0\cite{YF07} & 2.6 & 2.0 & 0.461 & 0.080 & 6.95 & 1.04 & 0.283 \\
d0.2-1.5\cite{YF07} & 2.6 & 1.5 & 0.463 & 0.081 & 2.62 & 0.99 & 0.240 \\
d0.7-1.5\cite{YF07} & 2.3 & 1.5 & 0.482 & 0.091 & 10.0 & 1.01 & 0.216 \\
23p-1.2\cite{YF07} & 3.2 & 1.2 & 0.362 & 0.080 & 0.655 & 0.992 & 0.0066 \\
23p-1.6\cite{YF07} & 2.4 & 1.6 & 0.439 & 0.079 & 23.5 & 0.996 & 0.613 \\
CL-20\cite{CL04} & 1.6 & - & 0.156 & 0.542 & 4.03 & 1.13 & 0.10 \\
CL-25\cite{CL04} & 1.8 & - & 0.245 & 1.26 & 2.19 & 2.44 & 0.10 \\
\hline

\end{tabular}
\caption{Yields for a range of models of a roughly 20-25\,M$_\odot$ star 
by different groups.  Where given in the literature, we include remnant 
mass and explosion energy.  Except for models 23e-series\cite{YF07}, 
the models all use piston explosions.  This is is evident from the 
small remnant masses for a given explosion (this can not be reproduced 
in a real explosion calculation).  Note that for a given explosion energy 
and remnant mass, we get considerable scatter in the yield (more than 
an order of magnitude).  Most of the difference is caused by those 
results using piston explosions and those using the more realistic 
models (which include fallback).}
\end{table}

\section{Comparing Nuclear Yields}

We have now discussed in detail the differences between the methods
used to simulate a explosions for supernova nucleosynthesis.  Table 1
shows the yields for models in the 20-25\,M$_\odot$ range by 3
different groups~\cite{FY07,WW95,CL04}.  The stars are all evolved
with an initial metallicity at solar, but prescriptions for winds vary
somewhat and each group uses its own method to drive explosions.  The
first difference between the models can be seen in the remnant masses.
Note that the WW models both predict remnant masses below
1.5\,M$_\odot$ for explosion energies of roughly
$2\times10^{51}$\,erg.  The CL remnants are also small.  But at the
same energy, the 23e-series produces a 2.6M$_\odot$ remnant.

The difference between these remnant masses is entirely an artifact of
our method of artificially induced explosions.  Different methods
produce very different amounts of fallback.  To better understand the
fallback, let's briefly review its history.  The idea of fallback was
first brought up by Colgate~\cite{Col71} to overcome nucleosynthesis
issues arising from the supernova ejection of neutron rich material
produced in stellar cores \cite{Arn71, Young06}.  Colgate argued that
the inner layers of the ejected material would deposit its energy to
the stellar material above it, ultimately reducing its energy below
that needed to escape the neutron star, and it would fall back onto
the neutron star.  In such a scenario, one would expect the inner
material to fall back quickly (within the first few to ten seconds).
It was argued that this material (the neutron rich material from the
initial explosion) would accrete onto the neutron star, alleviating
any nucleosynthesis issues.

Piston models for explosions misled many scientists on the issue of
fallback and the supernova field in general.  By artificially
preventing fallback, piston modelers became concerned with the ejecta
of neutron rich material (recall, this is why Colgate first thought
about fallback in the first place).  Supernova modelers have worked
extensively to try to reset the electron fraction and nucleosynthesis
modelers put in knobs to reset the electron fraction and move out the
mass cut.  The Colgate idea of fallback was all but forgotten.  With
more modern, energy-injected explosion models, fallback occurs
(renewing Colgate's original idea) and removes issues with neutron
rich ejecta.  This makes it easier for explosion models to match
compact remnant mass measurements~\cite{FK01} and may even explain the
r-process~\cite{FHHT06}.

Table 1 also shows the yields for some key elements from these models.  
There is a lot of scatter in these models, so it is difficult to pick 
out any specific trend, but note that some elements (e.g. $^{45}$Sc) 
are overproduced by piston models by more than a factor of 10.  Also, 
the ratio of $^{44}$Ti to $^{56}$Ni can be an order of 
magnitude higher in some energy-driven explosions (making it easier 
to explain the supernova that produced Cassiopeia A).  Not until 
we model a full suite of models will we truly understand the extent 
of the errors introduced by piston-driven models.

For our in-progress NuGrid calculations, we use a constant entropy 
injection process.  This is the closest match to the convection-enhanced 
explosion mechanism.  When the shock moves beyond 1000\,km, we stop the 
energy injection (which due to the entropy increase process starts to 
decrease as the density lowers anyway).  This still leaves 2 parameters:  
total energy injection and rate at which the energy is injected.  The 
rate has been studied at some level~\cite{YF07} and it can lead to 
order of magnitude differences in the yield.  Fortunately, for a given 
explosion energy, we can constrain the delay time~\cite{Fry06}, so 
we believe we can fix this parameter somewhat, limiting its errors.

\end{document}